\documentclass[a4paper,fleqn,usenatbib]{mnras}

\usepackage[T1]{fontenc}
\usepackage{ae,aecompl}

\usepackage{graphicx}	
\usepackage{amsmath}	
\usepackage{amssymb}	
\usepackage{multicol}        
\usepackage{bm}		
\usepackage{color}
\usepackage{xcolor}
\usepackage{soul}         
\usepackage{tabularx}
\newcommand\setrow[1]{\gdef\rowmac{#1}#1\ignorespaces}
\newcommand\clearrow{\global\let\rowmac\relax}
\clearrow

\title[Rapidly-oscillating {\it TESS} stars]
{Rapidly-oscillating {\it TESS} A--F main sequence stars:\\
are the roAp stars a distinct class?}
\author[L.A. Balona]{L. A. Balona\\
South African Astronomical Observatory, P.O. Box 9, Observatory, Cape
Town, South Africa}

\begin{document}

\date{Accepted .... Received ...}

\pagerange{\pageref{firstpage}--\pageref{lastpage}} \pubyear{2011}

\maketitle

\label{firstpage}

\begin{abstract}
From sector 1--44 {\em TESS} observations, 19 new roAp stars, 103 ostensibly 
non-peculiar stars with roAp-like frequencies and 617 $\delta$~Scuti stars 
with independent frequencies typical of roAp stars were found. 
Examination of all chemically peculiar stars observed by {\em TESS} resulted
in the discovery of 199 Ap stars which pulsate as $\delta$~Sct or $\gamma$~Dor 
variables.  The fraction of pulsating Ap stars is the same as the fraction
of pulsating chemically normal stars.  There is no distinct separation in 
frequency or radial order between chemically peculiar $\delta$~Sct stars and 
roAp stars.  In fact, all the features which originally distinguished roAp
from $\delta$~Sct stars in the past have disappeared.  There is no reason to
assume that the high frequencies in roAp stars are driven by a different
mechanism from the high frequencies in chemically normal stars.  However,
chemically peculiar stars are far more likely to pulsate with high
frequencies.   The term ``roAp'' should be dropped: all roAp stars are
normal $\delta$~Scuti stars.   
\end{abstract}


\section{Introduction}

The peculiar Ap and Fp stars have strong, approximately dipolar, kilogauss 
global magnetic fields with axes which are generally tilted with respect to the
rotation axes.  They have inhomogeneous surface abundances and brightness, 
leading to rotational light modulation (the oblique rotator model 
\citealt{Stibbs1950}).  Their spectra have unusually strong lines of some 
ionized metals (e.g., Sr, Cr) and rare earth elements.  The chemical 
peculiarities, which are confined to the outer layers, are thought to be a 
result of radiative acceleration, gravitational settling and diffusion.

The rapidly-oscillating Ap (roAp) stars are Ap/Fp stars which pulsate in a 
limited high-frequency range, typically above 60\,d$^{-1}$ \citep{Kurtz1978b,
Kurtz1982}.  When roAp stars were first discovered, the highest frequencies 
then detected among the $\delta$~Scuti stars were around 50\,d$^{-1}$.  These
frequencies could be understood by the operation of the opacity driven
$\kappa$~mechanism in the He\textsc{ii} partial ionization zone \citep{Cox1963}.  The 
higher frequencies seen in roAp stars, however, could not be reproduced by 
models and seemed to require a new driving mechanism.  For this reason, it 
seemed appropriate to make a distinction between $\delta$~Sct and roAp stars.

An additional reason why roAp stars were thought to be distinct from 
$\delta$~Sct stars is that high frequencies only seemed to occur among the 
chemically peculiar Ap stars.  Furthermore, while $\delta$~Sct stars 
pulsate in a broad range of frequencies, the roAp stars pulsate with just a 
single high frequency or else with multiple closely-spaced high frequencies.
Another important difference is that in some roAp stars the frequency peaks 
are split by an exact multiple of the rotation frequency, something which is 
never seen in $\delta$~Sct stars.  This can be understood by the oblique
pulsator model \citep{Kurtz1982, Kurtz1986}, in which the pulsation axis is
tilted with respect to the rotational axis.

In conventional models, pulsational driving due to the $\kappa$ mechanism 
acting in the H/He\,\textsc{i} partial ionization zone is almost enough to overcome 
damping in the rest of the star.  This led \citet{Balmforth2001} to suggest 
that suppression of  convection by a vertical magnetic field might be 
sufficient to reduce damping, resulting in driving of high-frequency pulsations
confined to the region around the magnetic poles (see also \citealt{Cunha2002},
\citealt{Cunha2013}).  This new model seem to finally solve the problem of
pulsational driving in roAp stars and explained why a strong magnetic field
is required for the pulsations to occur.  Since pulsational driving occurs
only at the poles, the model neatly justifies why only axisymmetric modes
are seen and the oblique pulsator model. 

Further attempts to understand high-frequency pulsations in $\delta$~Sct
stars were made by \citet{Antoci2014} who highlighted the role of turbulent
pressure in the envelope convective zone.  It was found that high frequencies 
could be excited in certain regions of the H--R diagram.  Turbulent pressure
is also considered as a possible mechanism in driving pulsations in Am stars
\citep{Smalley2017}.  The most comprehensive recent modelling of pulsations
in $\delta$~Sct stars is that of \citet{Xiong2016}, which also involves
turbulent pressure.  The models predict high frequencies in a limited
region of the H--R diagram.

Observations from the {\em Kepler} mission were the first to challenge the
perception that pulsations in the $\delta$~Sct stars are fully understood. 
The models are unable to reproduce the low frequencies (i.e. below 5\,d$^{-1}$) 
which are ubiquitous in $\delta$~Sct stars even at the highest effective 
temperatures \citep{Balona2014a}.  Neither can they reproduce the fact that 
both $\gamma$~Dor and $\delta$~Sct stars co-exist in the same effective 
temperature and luminosity region \citep{Balona2018c}.  Another problem is 
that less than half of the stars in the $\delta$~Sct instability region seem 
to pulsate \citep{Balona2018c}.  

{\em TESS} observations have also shown that high frequencies are not unique 
to roAp stars.  Some $\delta$~Sct stars have independent frequency peaks
extending into the roAp frequency range \citep{Balona2019b,Bedding2020}.
In this paper we present a list of 617 $\delta$~Sct stars with independent 
frequencies higher than 60\,d$^{-1}$. These are chemically normal stars, 
showing that high frequencies are not confined to chemically peculiar stars.  
This invalidates the need for a special driving mechanism involving strong 
magnetic fields and suppression of convection at the poles 
\citep{Balmforth2001}.

Furthermore, many ostensibly normal A-F stars pulsate in isolated high 
frequencies or tight high-frequency groups just as in roAp stars 
\citep{Cunha2019, Balona2019b}.  Many of these have turned out to be chemically
peculiar \citep{Holdsworth2021a}, but this may not be true for all stars.  In 
this paper we present more than 100 additional chemically normal stars with
frequencies and frequency distributions  that cannot be distinguished from 
roAp stars.

In this paper we also present the results of a survey of 1978 {\em TESS} Ap 
stars and the discovery that 199 are $\delta$~Sct or $\gamma$~Dor stars.  The 
idea that Ap stars do not pulsate is not correct.  These new discoveries
show that all the characteristics separating roAp stars from $\delta$~Sct 
stars are no longer applicable.  It is clear that whatever mechanism drives 
high frequencies in normal stars must surely drive the high frequencies in 
``roAp'' stars as well.  The equally-spaced frequencies seen in 
some roAp stars are not seen in normal $\delta$~Sct stars.  This is due to the 
action of a strong magnetic field on the atmospheric structure and does not 
require a separate pulsation mechanism.

The aim of this paper is to present new discoveries of ``roAp'' stars and
non-peculiar roAp-like pulsators as well as a survey of pulsation in
chemically peculiar stars.  Using these results, it is argued that a
suitable definition of the roAp class does not exist.  Furthermore, it is
argued that there is no criterion, other than a purely arbitrary frequency
criterion, which can distinguish roAp stars from chemically peculiar or
chemically normal $\delta$~Sct stars.  It is concluded that the roAp 
classification is no longer a helpful guide to understanding pulsations in A-F 
stars and that the name ``roAp'' is misleading and should be abandoned.

\section{The data and stellar parameters}

The {\em TESS} photometric survey consists of continuous wide-band photometry 
of 13 sectors per celestial hemisphere.  Each sector is observed continuously
for about 27\,days with 2-min cadence.  Stars near the ecliptic equator are 
only observed for one sector.  Stars in the circular regions nearer the
poles overlap so that, at the ecliptic poles, stars are observed continuously
for about 100\,days. The light curves are corrected for time-correlated 
instrumental signatures using pre-search data conditioning (PDC, 
\citealt{Jenkins2016}).   The data used here are the full PDC light curves
from sectors 1--44.

It should be noted that the {\em TESS} pixel size is  $21 \times 21$\,arcsec, 
which is roughly the size of the aperture used in ground-based photoelectric 
photometry. There is a possibility of contamination due to other stars in the 
same aperture.  However, the stars discussed here are bright, mostly in the 
range 8--10\,mag.  The chances of contamination by another star of comparable 
brightness is quite remote.  Contamination by a binary companion of comparable 
luminosity would be impossible to verify.

The most important step in any survey of this nature is to determine the
variability type of each star.  Automated classification is a relatively
simple task, but without visual examination of the light curves and
periodograms, there is no possibility of discovering anything new.  For
example, rotational modulation and flares in A stars  would not have been 
noticed \citep{Balona2013c}.  For this reason, visual inspection of the data 
is essential for further progress.

Using the {\tt SIMBAD} database \citep{Wenger2000}, a master catalogue of about 
684\,000 stars brighter than magnitude 12.5 and with spectral types earlier than 
G0 was created.  The catalogue includes the variability class if available,
the effective temperature, luminosity, rotation period, spectral type and
other useful parameters.  The spectral type is important for ascertaining the 
variability class. It is also particularly useful because photometric effective
temperature estimates are unreliable for B stars unless measurements in the U 
band are used (which is seldom the case).  For example, the effective 
temperatures of B stars listed in the {\em Kepler} or {\em TESS} catalogues are
not reliable.  The master catalogue is not restricted to stars observed by 
{\em Kepler} or {\em TESS}.

As  each {\it TESS} sector is released, stars in the bulk download file are 
matched with stars in the master catalogue and downloaded.  The light curve 
from the FITS file is extracted automatically and stored in a database.  If the
star is already in the database, the additional data is appended to the 
existing light curve.  Periodograms and extracted frequencies are calculated 
and form part of the database.  Usually there are about 1000--2000
unclassified stars for each {\em TESS} release.  Software tools (developed by 
the author) allow the periodogram and light curve to be displayed.  Tools for 
smoothing the light curve, identifying harmonics and other useful functions 
are also included.

By visual inspection of the light curve and periodogram, the author has 
assigned a variability class to each star.  In the few cases where a 
variability class is already known, there is good agreement. The task 
of visually classifying these stars typically takes two days.  Changes are 
sometimes made to previously classified stars as more data become available.  
After each data release, the master catalogue is updated with the new 
classifications.  The author has classified over 100\,000 stars hotter than 
about 6000\,K observed by {\em TESS} and {\em Kepler} in this way.

The variability classification follows that of the {\em General Catalogue of 
Variable Stars} (GCVS, \citealt{Samus2017}).  A/F stars with rotational light 
modulation due to chemical abundance spots (Ap stars) are classified as 
$\alpha^2$~CVn (ACV) variables.  Among the B stars these are known as SX~Ari 
(SXARI) variables.  

A new ROT class has been added to describe any star, not known to be 
chemically peculiar, in which the variability is suspected to be due to 
rotation.  The criterion for ROT requires the presence of just a single,
isolated frequency peak below 4\,d$^{-1}$ or a peak and its harmonic.  It
is not always possible to distinguish rotation from binary effects, but
to minimize such contamination, the ROT class is restricted to amplitudes 
below 10\,millimags. The $\beta$~Cep, SPB, Maia, $\delta$~Sct and
$\gamma$~Dor variables are typically multiperiodic and easily distinguished 
from the ROT stars.

A question arises as to what variability type must be assigned to a star
which would be classified as roAp on the basis of the frequencies, but where
the spectral type does not indicate chemical peculiarity.  One could simply
classify the star as a $\delta$~Sct.  But what if the spectral type is
incorrect and the star is indeed an Ap star?  In that case one would lose
the opportunity of detecting a new roAp variable.  In these circumstances
it is obviously important to make a note of these stars in some way.  The
solution that is adopted here is to label such stars as ``roA'' (ROA).  This 
does not imply a new class of stellar variability, of course.

A separate catalogue was created containing individual $T_{\rm eff}$ 
measurements and literature references for as many as possible of the 
684\,000 stars in the master catalogue. This effective temperature catalogue 
presently contains over 185\,000 individual $T_{\rm eff}$ measurements of 
about 101\,500 stars, over and above the effective temperatures from the 
{\it Kepler} and  {\it TESS} catalogues. The {\tt PASTEL} catalogue 
\citep{Soubiran2016} provided a very useful starting point for this compilation.
Subsequent entries were made by searching the {\tt SIMBAD} database. 

For each entry in the effective temperature catalogue, the method used to 
derive $T_{\rm eff}$ was noted and assigned an index, $t$.  If $T_{\rm eff}$ 
is derived from fitting a model atmosphere to the stellar spectrum, $t = 1$.  
If it is estimated from Str\"{o}mgren, Geneva or other narrow-band photometry, 
$t = 2$.  If $T_{\rm eff}$ is from the spectral energy distribution or similar 
method, $t = 3$.  If it is from the {\it Kepler} or {\it TESS} catalogues 
$t = 4$.  If it is estimated from the spectral type, $t = 5$.  The adopted 
value of $T_{\rm eff}$ is the average of all the values with lowest $t$.  
Measurements with higher $t$ values are ignored.  From time to time, the master
catalogue is updated with the adopted $T_{\rm eff}$.  

Photometric effective temperatures for Ap stars are unreliable because of line 
blanketing due to the spectral peculiarities.  For this reason, it is best
to use spectroscopic measurements ($t=1$) whenever possible.  Table\,\ref{teff}
lists all individual $T_{\rm eff}$ values and references used to determine 
the adopted $T_{\rm eff}$ for stars listed in the Appendix.  The table 
contains 574 entries for all 443 stars mentioned in this paper.  For these 
stars, a thorough literature search was made to ensure that all available 
measurements are included.

Spectral types are mostly from the catalogue of \citet{Skiff2014} supplemented 
by later publications when required.  The calibration of \citet{Pecaut2013} 
was used to estimate $T_{\rm eff}$ as a function of spectral type for those
stars where the spectral type is the only available method of estimating
$T_{\rm eff}$ ($t = 5$).

The stellar luminosity listed in the tables was estimated from {\it Gaia} EDR3 
parallaxes  \citep{Gaia2016, Gaia2018} in conjunction with reddening obtained 
from a three-dimensional map by \citet{Gontcharov2017} using the bolometric correction 
calibration by \citet{Pecaut2013}. From the error in the {\it Gaia} EDR3 
parallax, the typical standard deviation in $\log(L/L_\odot)$ is estimated to 
be about 0.05\,dex, allowing for standard deviations of 0.01\,mag in the 
apparent magnitude, 0.10\,mag in visual extinction and 0.02\,mag in the 
bolometric correction in addition to the parallax error.  However, comparison
between EDR3 and DR2 releases of the {\em Gaia} parallaxes suggests that
the real error in luminosity is probably closer to 0.07\,dex.

\begin{table}
\begin{center}
\caption{Catalogue of effective temperatures, $T_{\rm eff}$, index, $t$, and
literature reference. The full table is available in electronic form.} 
\label{teff}
\begin{tabular}{rrrr}
\hline
\multicolumn{1}{c}{TIC}              & 
\multicolumn{1}{c}{$t$}              & 
\multicolumn{1}{c}{$T_{\rm eff}$}    & 
\multicolumn{1}{c}{Ref}              \\
\hline
  2849758 &  4 &  9033 & {\tt 2018AJ....156..102S} \\
  3373254 &  5 &  7790 & {\tt 2013ApJS..208....9P} \\
  3542929 &  5 &  8500 & {\tt 2013ApJS..208....9P} \\
  3814749 &  2 &  7861 & {\tt 2020A\&A...638A..76Q} \\
  3814749 &  2 &  8029 & {\tt 2019A\&A...628A..94A} \\
  4463975 &  1 &  7629 & {\tt 2020ApJS..247...28H} \\
  5370150 &  2 &  6927 & {\tt 2019A\&A...628A..94A} \\
\hline
\end{tabular}
\end{center}
\end{table}

\section{Comparison between chemically peculiar $\delta$~Sct and roAp stars}

We need to ask what characteristics are traditionally used in classifying a 
star as roAp. The general view is that roAp stars are simply Ap stars with high 
frequencies, but there is no consensus on the meaning of ``high frequencies''.
The lowest frequency in the group of recognized roAp stars (as listed in 
\citealt{Smalley2015}) is 61\,d$^{-1}$.  It is reasonable to suppose,
then, that ``high frequencies'' means any frequency higher than about
60\,d$^{-1}$. 

Definition of a variability group should be based on physical considerations
and not arbitrary choices.  However, since no other definition has been
proposed, it will be assumed that a star is roAp only if the pulsation
frequency is higher than the hard limit of 60\,d$^{-1}$.  Low-frequency
peaks which can be attributed to rotation are often detected as well.
Lower frequencies typical of $\gamma$~Dor or $\delta$~Sct stars may co-exist
with the high frequencies, in which case the star is given a hybrid
classification, e.g. DSCT+ROAP.

Note that recently \citet{Holdsworth2021a} admitted TIC\,356088697, which has 
a single pulsation mode at 55.8\,d$^{-1}$, to the list of roAp stars.  There 
is no real reason to query the lowering of the frequency limit, because there 
is no theory which provides a guide on the  supposed frequency range in which 
roAp stars may pulsate.  In fact, one might as well use any limit that
feels right.  This is clearly not a satisfactory situation.

Table\,\ref{known} lists all known roAp stars, even those not observed by 
{\em TESS}.  This is essentially the list in \citet{Smalley2015} supplemented 
by {\em TESS} discoveries by \citet{Cunha2019} and \citet{Balona2019b} and,
most recently, by \citet{Holdsworth2021a}. The sources of $T_{\rm eff}$ for 
these stars are listed in Table\,\ref{teff}.  The rotation periods in this and 
other tables  are either those in the literature or updated by the author 
using the {\em TESS} light curves.  The characteristic frequency, $\nu$, is the 
frequency of highest amplitude above 60\,d$^{-1}$.

The Am (metallic-lined) stars are characterized by an under-abundance of Ca 
(and/or Sc) and/or an over-abundance of Fe and the iron-group elements. Unlike 
the Ap stars, the abundances of rare earth elements are normal and a global 
magnetic field is absent or very weak.  The high metal abundance 
is thought to be a result of diffusion in the absence of a magnetic field 
\citep{Michaud1976}.  As mentioned by \citet{Holdsworth2014a}, it is 
difficult to distinguish between Am and Ap stars at classification dispersion.  

For this reason, and because Am stars are not known to pulsate with  high 
frequencies, stars classified as Am are included in the tables as possibly 
mis-classified roAp stars.  Rotational variables among Am stars are assigned 
the ROT class (not ACV).  Some stars which are not known to be Ap, but which 
have been accepted in the literature as being roAp are also listed.

\begin{figure}
\centering
\includegraphics[]{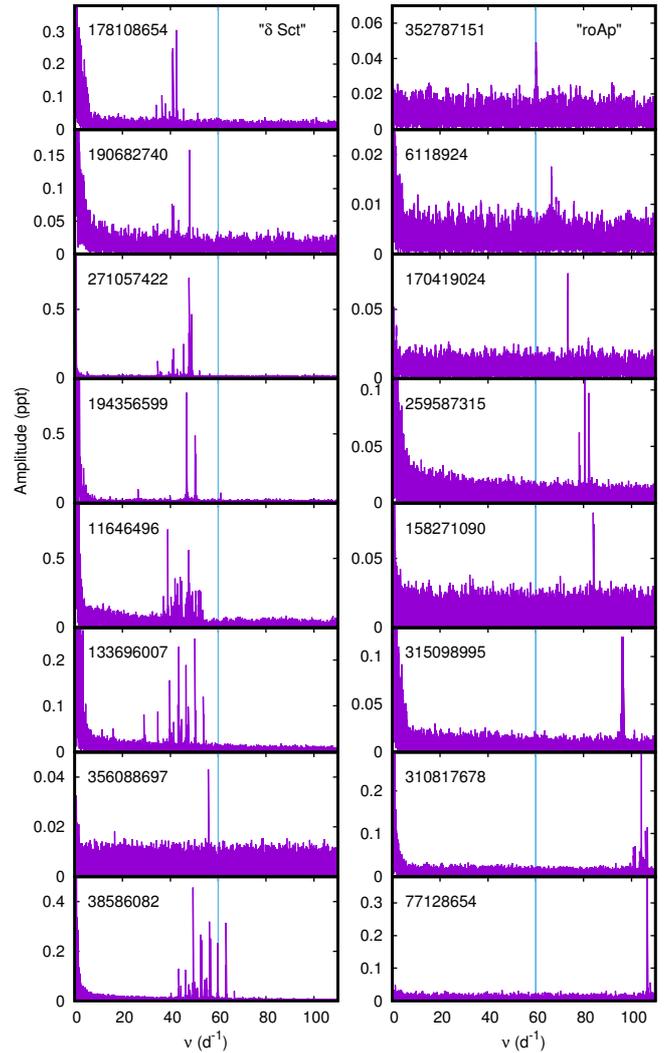}
\caption{The periodograms of some Ap stars arranged in order of increasing 
frequency.  The left panel shows what may be classified as $\delta$~Scuti 
variables while the right panel are all known roAp stars.  As can be seen, 
there is no distinct separation between the two classes of Ap pulsating stars
except for the artificial boundary at 60\,d$^{-1}$ (vertical line).}
\label{ap}
\end{figure}

All known chemically peculiar stars observed by {\em TESS} were examined for
pulsational variability.  Quite a large number of $\delta$~Sct and
$\gamma$~Dor Ap stars were found (Table\,\ref{droap}).  These will be
discussed separately below.

Fig.\,\ref{ap} shows the periodograms of a number of Ap $\delta$~Sct
stars from Table\,\ref{droap} as well as roAp stars from Table\,\ref{known}.  
We note that there is no obvious frequency separation between stars 
which are classified as roAp, i.e. those with frequencies higher than 
60\,d$^{-1}$ (right panel) and those of slightly lower frequencies (left 
panel), which we must presumably regard as Ap $\delta$~Sct stars.  Note that 
the stars shown in the figure have single or isolated multiple high 
frequencies typical of roAp stars. The situation where a variability group
is defined on the basis of an arbitrary frequency limit, but where no basis
for such a limit exists, is clearly not satisfactory.

The main characteristic used to define roAp stars - that of high frequencies
- can no longer be used.  There is no frequency which marks an obvious lower
limit to pulsations in Ap stars.  Neither is there a discernible frequency
gap between $\delta$~Sct and roAp classification.  The argument that
chemically peculiar stars do not pulsate cannot be used to justify the
roAp class either.

\section{New roAp and roAp-like stars}

Table\,\ref{new} lists previously unreported {\em TESS} stars with effective 
temperatures $T_{\rm eff} > 6000$\,K in which at least one significant, 
independent frequency peak higher than 60\,d$^{-1}$ is present.  The test for 
significance is based on the criterion that the peak amplitude must exceed the
mean local noise level by a certain factor.  It is not possible to directly 
calculate the false alarm probability in this way.  The generally used 
criterion of S/N $> 4$ \citep{Breger1993} is probably too optimistic 
\citep{Baran2015, Baran2021, Bowman2021}.  Here we use the criterion 
S/N$>4.7$, but of course there will always be uncertainty at low values of 
S/N.

Table\,\ref{new} lists 19 known Ap stars with isolated high-frequency peaks 
or group of peaks higher than 60\,d$^{-1}$ which have not been previously
reported.  These would certainly be accepted as new stars of the roAp class.
However, the majority of new pulsating stars in Table\,\ref{new} are chemically
normal.  From this perspective, they are just $\delta$~Sct stars unless further
study shows that they are chemically peculiar.  Labeling these roAp-like stars
as ``roAp'' is not correct as there is no evidence that they are chemically 
peculiar.  Classifying them as $\delta$~Sct would also not be correct because 
it is possible that some may turn out to be chemically peculiar.  For this 
reason, the temporary label ``roA'' has been assigned to them.  This does not 
signify a new class of variable star, but is simply a label to identify stars 
which cannot be properly classified until chemical peculiarity has been 
satisfactorily established or ruled out.

\begin{table}
\begin{center}
\caption{Bright ROA stars discovered from {\em TESS} photometry.}
\label{roa}
\resizebox{\columnwidth}{!}{
\begin{tabular}{rllrl}
\hline
\multicolumn{1}{c}{TIC}              & 
\multicolumn{1}{c}{Name}             & 
\multicolumn{1}{c}{Var. Type}        & 
\multicolumn{1}{c}{$V$}              &
\multicolumn{1}{c}{Sp. Type}         \\
\hline
 313261813 & HD 119476  & GDOR+ROT+ROA &  5.85 &  A1.5V   \\
 398476730 & HD 104125  & ROT+ROA      &  6.76 &  A2V     \\
 464807201 & HD 29839   & ROA+ROT      &  7.10 &  A1V     \\
 259017938 & HD 210684  & ROT+ROA+r    &  7.37 &  F0      \\
\hline
\end{tabular}
}
\end{center}
\end{table}

\begin{figure}
\centering
\includegraphics[]{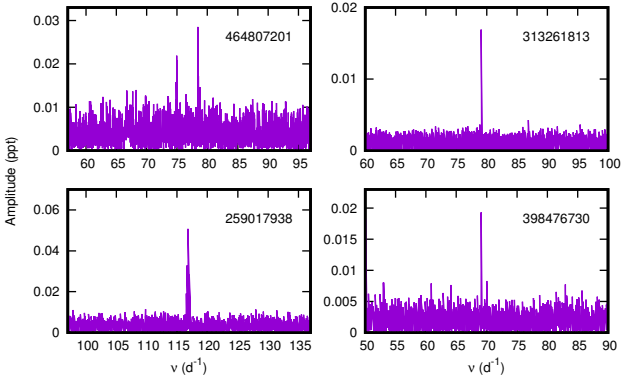}
\caption{Examples of bright stars, not known to be chemically peculiar, with 
high roAp-like frequencies.}
\label{bright}
\end{figure}

It would be expected that chemical peculiarities  would be most easily 
detected in bright stars, yet the spectral classifications of many of the 
brightest roA stars do not mention chemical peculiarity.  Four of these bright 
roA stars are listed in Table\,\ref{roa} and their periodograms shown in 
Fig.\,\ref{bright}.  As can be seen, these would certainly be classified as 
roAp stars. HD\,119476 (TIC\,313261813), for example, has been given
spectral types in the range B9V--A2V by seven different authors 
\citep{Skiff2014} and in no case was chemical peculiarity identified. 

\begin{figure}
\centering
\includegraphics[]{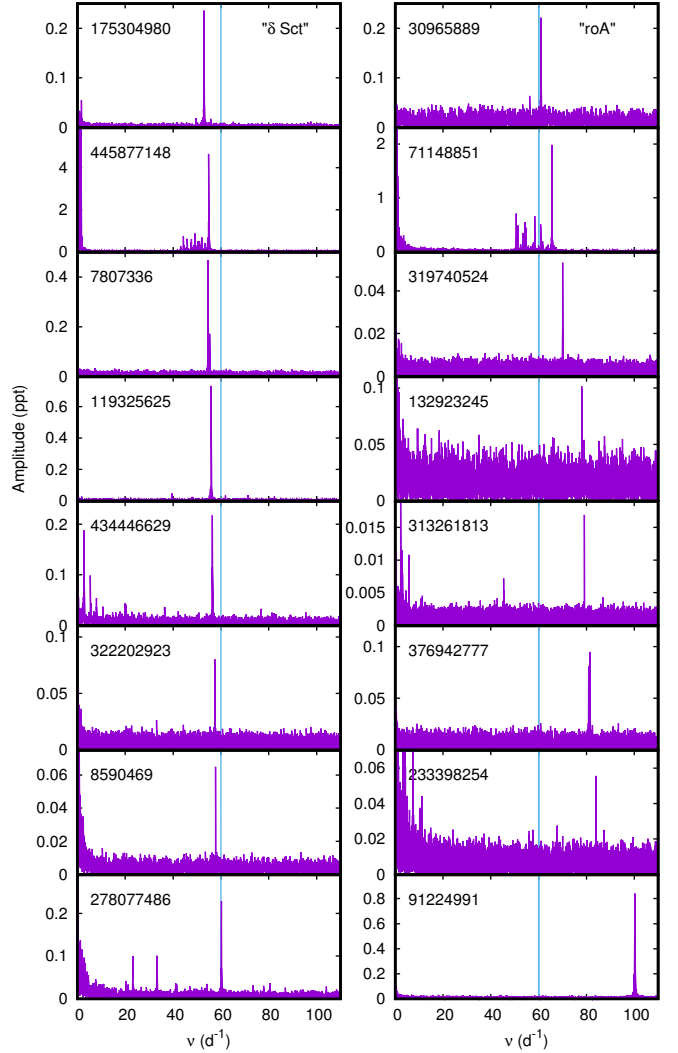}
\caption{The periodograms of some ostensibly chemically normal stars arranged 
in order of increasing frequency. The vertical line is the adopted limit of 
60\,$^{-1}$.  By analogy with the roAp stars, stars with at least one
frequency higher than this limit are called ``roA'' stars, even though this
distinction is purely arbitrary.}
\label{noap}
\end{figure}

Fig.\,\ref{noap} shows periodograms of some ostensibly chemically normal
roA stars arranged in order of increasing frequencies.  We see that, as in the
roAp stars, there is no separation between the roA and normal $\delta$~Scuti 
stars.  What this figure also shows is that isolated frequency peaks, which
are typical of roAp stars, are quite common among $\delta$~Sct stars as well  
(see \citealt{Balona2021c}).

\begin{figure}
\centering
\includegraphics[]{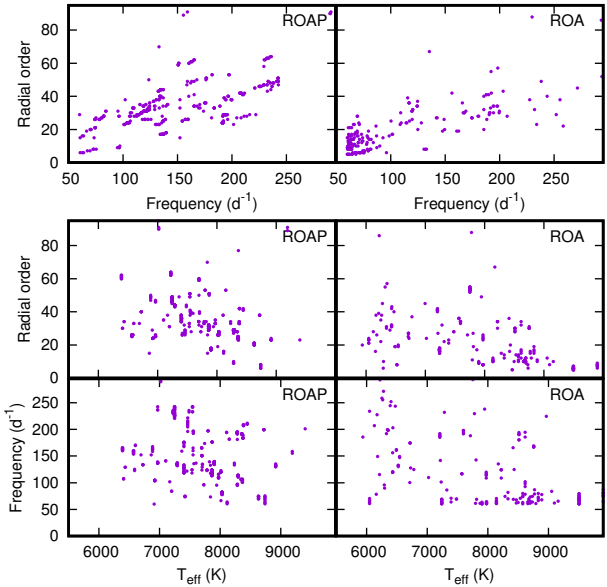}
\caption{Top panel: the radial order, $n$, as a function of pulsation
frequency.  Middle  panel: the radial order, $n$, as a function of effective 
temperature.  Bottom panel: pulsation frequency as a function of effective 
temperature.  The left panel shows stars recognised as roAp
(Table\,\ref{known}).  The right panel shows stars which have the same
photometric characteristics as roAp stars but are not known to be chemically
peculiar.  These are presumably normal $\delta$~Sct stars, but here they are
called roA stars because some might be misclassified Ap stars.}
\label{teffn}
\end{figure}

\section{Radial order as a possible discriminant}

It may perhaps be possible to discriminate between roAp stars and 
roAp-like $\delta$~Sct stars on the basis of pulsational radial order,
as attempted in \citet{Balona2019b}.  This is not likely to succeed because 
there is clearly a strong correlation between frequency and radial order.  
Furthermore, there is no physical basis why radial order should be 
important, but it is worth the  attempt.

The radial order, $n$, associated with pulsation frequency for known roAp
stars in Tables\,\ref{known} was derived from the stellar parameters and 
pulsation models by \citet{Dziembowski1977a}.  Fig.\,\ref{teffn} compares
the radial order calculated for roAp stars (left panels) and for roA stars
(right panels).  The top panel shows the correlation between pulsation 
frequency and radial order.  The scatter is due to stars in different
stages of evolution.  The middle panel shows $n$ as a function of 
effective temperature.  The bottom panel shows the pulsation 
frequency as a function effective temperature.

One might suppose, for example, that roAp pulsation might only occur above
a certain radial order number.  In that case the plot of radial order as a
function of $T_{\rm eff}$ should show a hard boundary.  As can be seen, 
this is not the case and the range in radial order is just as large as
the range in frequency for both roAp and roA stars.  Using the radial 
order offers no advantage over frequency as a means of distinguishing 
roAp/roA stars from  $\delta$~Sct stars.  

\begin{table*}
\begin{center}
\caption{List of 617 new $\delta$~Sct stars with independent high frequencies
exceeding 60\,d$^{-1}$. The columns are the same as in Table\,\ref{known}.
The full table is available in electronic form.}
\label{dsct}
\begin{tabular}{rllrrrrl}
\hline
\multicolumn{1}{c}{TIC}                         & 
\multicolumn{1}{l}{Name}                        & 
\multicolumn{1}{l}{Var Type}                    & 
\multicolumn{1}{c}{$\nu$}                       & 
\multicolumn{1}{c}{$T_{\rm eff}$}               & 
\multicolumn{1}{c}{$\log \tfrac{L}{L_\odot}$}   & 
\multicolumn{1}{c}{$P_{\rm rot}$}               & 
\multicolumn{1}{l}{Sp Type}                     \\
\multicolumn{1}{c}{}                            & 
\multicolumn{1}{c}{}                            & 
\multicolumn{1}{c}{}                            & 
\multicolumn{1}{c}{(d$^{-1}$)}                  & 
\multicolumn{1}{c}{(K)}                         & 
\multicolumn{1}{c}{(dex)}                       & 
\multicolumn{1}{c}{(d)}                           & 
\multicolumn{1}{c}{}                            \\
\hline
  1221946 & HD 30461        & DSCT+ROA        & 62.016  & 7671 & 1.03 &        & A2/3II   \\
  1404122 & HD 80750        & DSCT+ROA        & 60.152  & 8981 & 1.26 &        & A0V      \\
  2004993 & HD 34113        & DSCT+ROT+ROA    & 60.265  & 8885 & 1.31 &  7.143 & A0       \\
  3891160 & HD 99120        & DSCT+ROA        & 61.420  & 8614 & 1.21 &        & A1V      \\
  4154746 & HD 294001       & DSCT+ROA        & 61.945  & 8490 & 1.12 &        & A2       \\
  4202325 & HD 35221        & DSCT+ROA        & 71.942  & 8506 & 1.07 &        & A2       \\
  4250763 & HD 35318        & DSCT+ROA        & 78.658  & 9414 & 1.40 &        & A0       \\
\hline
\end{tabular}
\end{center}
\end{table*}

\begin{figure}
\centering
\includegraphics[]{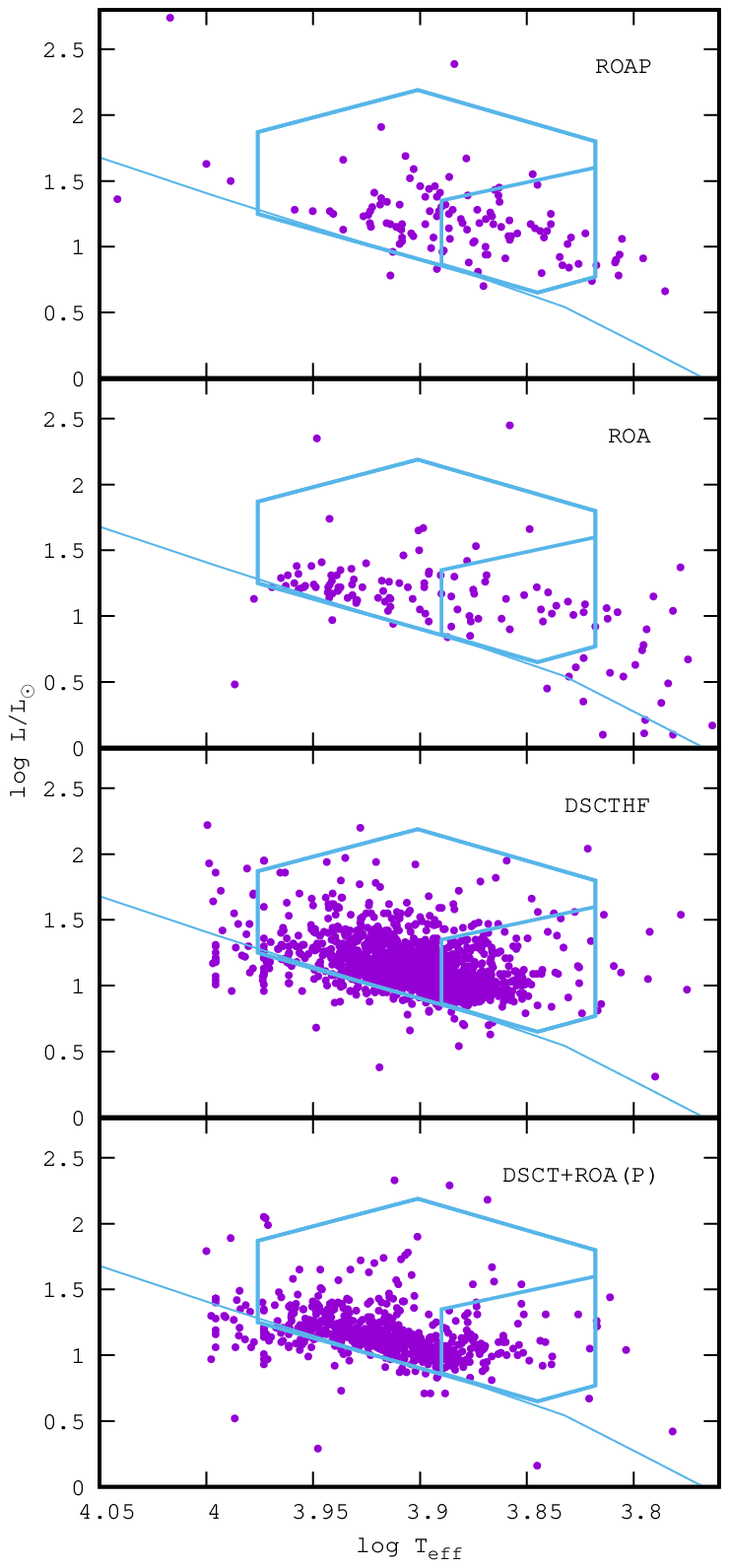}
\caption{The top two panels show the locations of the roAp and roA stars in
the theoretical H-R diagram.  The third panel shows $\delta$~Sct stars with 
frequency peaks $50 < \nu < 60$\,d$^{-1}$ (high-frequency $\delta$~Sct stars, 
DSCTHF).  The bottom panel shows $\delta$~Sct stars with at least one 
independent frequency $\nu > 60$\,d$^{-1}$.  The solid line is the zero-age 
main sequence for solar abundance models by \citet{Bertelli2008}.  The polygons
are the approximate areas where most $\delta$~Scuti and $\gamma$~Dor stars are 
found \citep{Balona2018c}.}
\label{hrdiag}
\end{figure}

\section{Hybrid $\delta$~Sct roAp and high-frequency $\delta$~Sct stars}

Among the recognised roAp stars in Table\,\ref{known}, there are 10
$\delta$~Sct+roAp hybrids.  A further three new $\delta$~Sct+roAp hybrids
are listed in Table\,\ref{new}.  It is, of course, possible that these stars 
are all multiple systems in which one star is $\delta$~Sct and the other a pure
roAp star. This cannot be ruled out, but appears rather contrived.  Recently, 
\citet{Murphy2020} have concluded that most of these $\delta$~Sct+roAp hybrids 
are not multiple stars.

There is a large number of chemically normal  $\delta$~Sct variables 
with high frequencies.  Whereas roAp and roA stars have only a few isolated 
high-frequency peaks, these high-frequency $\delta$~Sct stars have rich 
frequency spectra extending well into the roAp range.  Most of the
high frequencies are combinations and harmonics arising from non-linear 
interaction of low-frequency independent modes.  

Independent frequencies may be determined by looking at all possible 
combinations of parent frequencies, $\nu_1$ and $\nu_2$, such that 
$\nu = n_1\nu_1 + n_2\nu_2$, where $n_1$ and $n_2$ are positive or negative 
integers with $|n_1| \le |n_{\rm max}|$, $|n_2| \le |n_{\rm max}|$.  The
value of $n_{\rm max}$ should not be too large because for a sufficiently 
large value, almost any frequency can be described by a combination.  It
should also not be too low as to miss on a possible low-order interaction.
As a compromise, $n_{\rm max} = 7$ was chosen.  In deciding whether a 
combination, $\nu$, is significantly different from the observed frequency, 
$\nu_{\rm obs}$, the criterion $|\nu - \nu_{\rm obs}| < 3\sigma$ was used, 
where $\sigma$ is given by
            $$\sigma^2 = \sigma_{\rm obs}^2 + \sigma_1^2 + \sigma_2^2,$$
and $\sigma_{\rm obs}$ is the standard deviation of the observed
frequency.  The standard deviations of the parent frequencies are $\sigma_1$
and $\sigma_2$.

There are at least  617 $\delta$~Sct stars with independent high frequencies 
in the roAp range.  These are listed in Table\,\ref{dsct}.  To distinguish them 
from other high-frequency $\delta$~Sct stars, they are labeled as 
$\delta$~Sct+roA. This is not meant to be a new class of variable, but just a 
convenient label.

It is not clear why $\delta$~Sct+roA stars should be driven by a different 
mechanism or how such stars are to be differentiated from $\delta$~Sct+roAp 
hybrids.  In other words, chemically normal $\delta$~Sct stars with high 
frequencies can be considered to be in the same group as chemically peculiar 
$\delta$~Sct stars with high frequencies. There is no reason why the pulsation 
mechanism in these two groups should be different.

\section{Location of stars in the H--R diagram}

High-frequency stellar pulsations are to be found in stars with high mean
densities.  As Fig.\,\ref{hrdiag} shows, the $\delta$~Sct+roA or 
$\delta$~Sct+roAp stars are found close to the zero-age main sequence 
where the mean densities are the largest.  One may expect stars with lower
frequencies to be more evolved because they would be less dense and have
larger radii.  To show this, $\delta$~Sct with frequencies 
higher than 50\,d$^{-1}$ but not higher than 60\,d$^{-1}$ (labeled as
DSCTHF in the figure), are, on average, slightly more evolved, as expected.  

Most roAp and roA stars populate the cooler half of the $\delta$~Sct 
instability region.  There are a significant number of roA stars cooler than 
the red edge of the $\delta$~Sct instability strip defined by the {\em Kepler}
$\delta$~Sct stars.  The pulsations in these cool roA stars cannot be mistaken 
for solar-like oscillations because the typical Gaussian envelope of peak 
amplitudes is not seen.  Furthermore, the stars do not obey the scaling law 
for solar-like oscillations \citep{Kjeldsen1995a}.  However, there are 
indications that the $\delta$~Sct instability region defined by {\em TESS} 
stars is extended towards cooler stars compared to the {\em Kepler} 
instability region  shown in Fig.\,\ref{hrdiag}. This may be a result of 
metallicity differences between the {\em Kepler} and {\em TESS} stars.
Most likely these cool roA stars are simply $\delta$~Sct stars in which
isolated high frequencies in the roAp range are selected.

\section{A\texorpdfstring{\MakeLowercase{p}}{p} stars with $\delta$~Scuti pulsations}

\citet{Saio2005} performed a non-adiabatic analysis of axisymmetric nonradial 
pulsations in the presence of a dipole magnetic field for a model with 
$M = 1.9$\,$M_\odot$ in which convection was suppressed in the stellar
envelope.  It was found that $\delta$~Sct pulsations are damped if the polar 
field strength is larger than about 1\,kG.   However, high-order p modes with 
frequencies corresponding to roAp stars driven by the $\kappa$~mechanism 
in the H\textsc{i} ionization zone remain overstable, even in the presence of a strong 
magnetic field.  This analysis suggests that roAp pulsations should not 
co-exist with $\delta$~Sct pulsations in Ap stars unless the magnetic field 
strength is lower than about 1\,kG.  

More recent calculations by \citet{Murphy2020} indicate that the fundamental 
mode can be excited even in the presence of a magnetic field as strong as 4kG, 
but high radial order g modes typical of $\gamma$~Dor pulsations are strongly 
suppressed.  If this prediction is correct, there should not be many Ap stars
with low-frequency $\gamma$~Dor pulsations.  There are 199 Ap/Fp 
stars in which $\delta$~Sct or $\gamma$~Dor pulsations are present 
(Table\,\ref{droap}).  Most of the Ap $\delta$~Sct stars have multiple 
low-frequency peaks indicative of high radial order g modes, suggesting
a possible problem in the models used by \citet{Murphy2020}.

The prevailing view is that $\delta$~Sct pulsations are not found, or are very 
rare, among Ap stars.  As Table\,\ref{droap} shows, this is not the case. 
{\em Kepler} observations of TIC\,26418690 (KIC\,11296437), the star discussed 
by \citet{Murphy2020}, show that it is a $\delta$~Sct+roAp star, but the 
$\delta$~Sct peaks have too low an amplitude to be seen in {\em TESS} data.  
\citet{Murphy2020} argue that it is not a binary and that the $\delta$~Sct and 
roAp pulsations originate in the same star. 

It is interesting to note that the number of $\delta$~Sct+roAp stars relative 
to Ap $\delta$~Sct stars (13/199 or about 7\,percent), is about 
the same as the number of $\delta$~Sct+roA stars relative to non-Ap 
$\delta$~Sct stars (617/8370 or 7\,percent).

\begin{figure}
\centering
\includegraphics[]{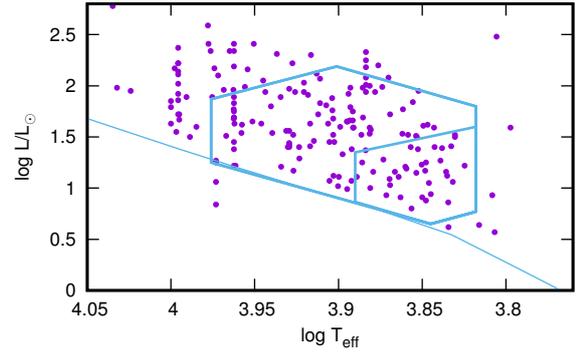}
\caption{Location of chemically peculiar $\delta$~Sct stars of
Table\,\ref{droap} in the theoretical H-R diagram.}
\label{hrap}
\end{figure}

To compare the relative number of pulsating Ap/Fp stars with those of normal
A/F stars, only main-sequence stars brighter than 10-th magnitude and with
$6000 < T_{\rm eff} < 9000$\,K were selected.  The magnitude limit
ensures that the sample is fairly complete.  With these restrictions, there
are 37348 stars observed by {\em TESS} of which 7104 are classified as 
$\delta$~Sct or $\gamma$~Dor.  Thus about 19\,percent of all main sequence 
stars in this temperature range are pulsating $\delta$~Sct/$\gamma$~Dor
variables.  

In the same effective temperature and magnitude ranges, there are
875 Ap/Fp stars observed by {\em TESS} of which 89 are $\delta$~Sct 
or $\gamma$~Dor stars.  Among Ap/Fp stars, about 10\,percent pulsate as 
$\delta$~Sct or $\gamma$~Dor. In the the same $T_{\rm eff}$ and magnitude 
ranges, there are 60 {\em TESS} roAp stars.  Therefore the fraction of
pulsating stars among the chemically peculiar stars is (89+60)/875 or about
17\,percent.  One may conclude that the fraction of pulsating Ap stars is
about the same as the fraction of pulsating chemically normal stars.  
However, there is a very clear tendency for higher frequencies among Ap stars.

Fig.\,\ref{hrap} shows the stars listed in Table\,\ref{droap} in the 
theoretical H-R diagram.  It is clear that the chemically peculiar 
$\delta$~Sct stars occupy the same instability region as chemically normal 
$\delta$~Sct stars.

\section{Critical frequencies}

It has been known for a long time that the pulsation frequencies in a 
significant number of roAp stars exceed the acoustic critical frequency,
$\nu_{\rm crit}$.  This is the frequency beyond which the pulsational
acoustic waves are no longer reflected in the atmosphere.  Modes with $\nu >
\nu_{\rm crit}$ manifest as running waves, leading to energy loss and
damping of pulsational driving.  

\begin{figure}
\centering
\includegraphics[]{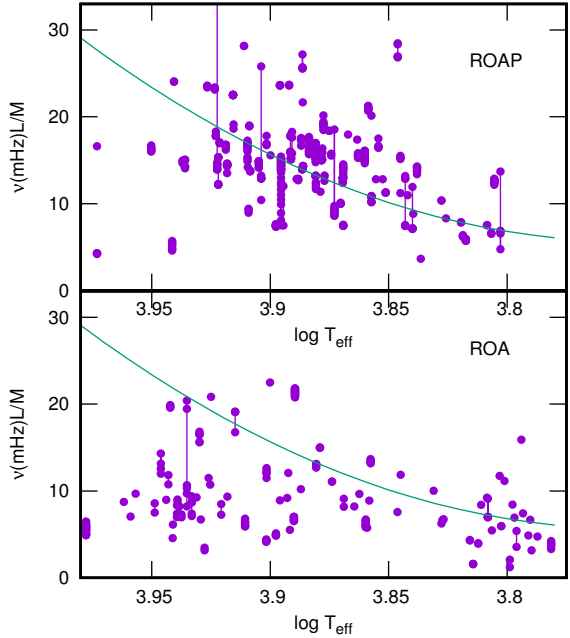}
\caption{The location of the roAp and roA stars in the $\log T_{\rm eff}$ -- 
$\nu L/M$ plane where $\nu$ is the frequency in mHz and $L$, $M$ is the 
luminosity and mass in solar units.  Different frequencies in 
the same star are connected by a vertical line (harmonics ignored).  The solid 
curve is the approximate acoustic critical frequency from models.}
\label{fcrit}
\end{figure}

As pointed out by \citet{Saio2014}, the critical frequency is best shown in
the $\log T_{\rm eff}$ -- $\nu L/M$ plane where $\nu$ is the frequency in mHz 
and $L$ and $M$ are the luminosity and mass in solar units.  Fig.\,\ref{fcrit} 
shows the location of the roAp and roA stars in this diagram.   The calculated 
critical frequency is adapted from \citet{Saio2014} and \citet{Audard1998}.
Note that the frequencies in about half the roAp stars exceed
the critical acoustic frequency.  However, for most of the roA stars the
pulsation frequency is below critical.  

It is not clear if this result is significant owing to uncertainties in the 
stellar parameters as well as uncertainties in the atmospheric models used to 
estimate $\nu_{\rm crit}$.  In any case, it is not clear whether the
atmospheric structure in highly magnetic stars is the same as in
non-magnetic stars.  Perhaps the difference can be reconciled in this way
and that $\nu_{\rm crit}$ is underestimated in the Ap stars.  If Ap stars do
have systematically higher critical frequencies than normal stars, it might
explain why high frequencies are more common among Ap stars.

\begin{figure}
\centering
\includegraphics[]{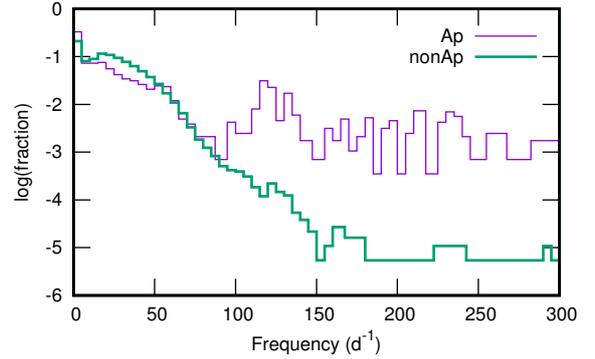}
\caption{The frequency distribution for $\delta$~Sct+$\gamma$~Dor+roA stars
(thick (green) line) and for Ap $\delta$~Sct+$\gamma$~Dor+roAp stars (thin
violet) line.  This is the fraction of independent frequencies within a 
frequency interval of 5\,d$^{-1}$. All frequency peaks with S/N > 4.7 are 
given the same weight (amplitude ignored).  }
\label{frqdis}
\end{figure}

\section{Conclusions}

It is generally believed that chemically peculiar Ap/Fp stars do not pulsate
as $\delta$~Sct or $\gamma$~Dor variables.  In this paper we have examined
1978 known Ap/Fp stars observed by {\em TESS} and found that not only do
some pulsate as roAp stars, but 199 pulsate as $\delta$~Sct or $\gamma$~Dor
stars (Table\,\ref{droap}).  Furthermore, as Fig.\,\ref{ap} shows, there is
no separation in frequency between these chemically peculiar $\delta$~Sct stars
and the roAp stars.  If this were known when the roAp stars were first 
discovered, there would have been no motivation for creating a separate roAp 
class at all.

The roAp class, in any case, has never been properly defined.  In this paper a 
hard frequency limit of 60\,d$^{-1}$ is used to separate $\delta$~Sct from
``roAp'' stars.  This is a purely arbitrary figure. No actual frequency or
radial order limit can be established.

The {\em TESS} observations have shown that isolated high frequencies
characteristic of roAp stars can be found in stars which have not been noted
as chemically peculiar.  In this paper, 103 new stars of this kind are
presented (Table\,\ref{new}), in addition to those previously discovered 
\citep{Cunha2019, Balona2019b}.  Some, but not all, of ostensibly
non-peculiar stars with isolated frequency peaks in the roAp range have been 
subsequently found to be chemically peculiar \citep{Holdsworth2021a}.

{\em TESS} observations of $\delta$~Sct stars show that many pulsate with 
frequencies in the roAp range.  The high frequencies in most of
these stars are due to combinations or harmonics.  An analysis of a large
number of $\delta$~Sct stars resulted in the discovery of at least 617
high-frequency $\delta$~Sct stars with independent pulsation frequencies in
the roAp range (Table\,\ref{dsct}).

All the reasons which motivated the creation of a new class of variable,
the roAp stars, have disappeared.  It is recommended that the term ``roAp''
be avoided and that these be considered as normal $\delta$~Sct stars. There is
no reason to suppose that the high frequencies in normal $\delta$~Sct stars
are driven by a different mechanism than in the ``roAp'' stars.  The need for 
the  mechanism described by \citet{Balmforth2001} falls away.

The strong magnetic field in Ap stars does, however, affect pulsations in
some stars.  This is evident in about one-third of the roAp stars in which
equidistant frequency peaks separated by the rotational frequency are seen.
It is likely that the strong magnetic field modifies the atmospheric structure 
so that the sound speed varies in conjunction with the magnetic field over the 
surface of the star.  Under some circumstances, the pulsational amplitude 
may vary significantly across the surface.  This will only be noticeable
in the highest frequencies where the pulsational wavelength is comparable with
the atmospheric scale height.  This explains, in a natural way, why the 
pulsation axis appears to be aligned with the magnetic axis in some roAp 
stars (the oblique pulsator model, \citealt{Kurtz1982, Kurtz1986}).

It has been known for some time that the spectroscopic line profile 
variability in roAp stars is characterized by blue-to-red moving features
(e.g. \citealt{Kochukhov2001b, Kochukhov2001a}).  This behaviour is common in 
many rotating nonradial pulsators and indicates non-axisymmetric pulsations.
However, this is inexplicable in the framework of the standard oblique 
pulsator model of slowly-rotating roAp stars which demands axisymmetric
modes.  Why only axisymmetric modes seem to be selected in not known.  
\citep{Kochukhov2007b} suggests that the modes can still be axisymmetric
if the line width varies with height in the atmosphere.  In the context 
proposed here, where the pulsation amplitude varies with magnetic field 
strength because of changes in sound speed, there is no need to assume 
axisymmetric modes in the oblique pulsator model.  Naturally, the pulsation
will be magnetohydrodynamic in nature and not purely acoustic.

\citet{Saio2005} and \citet{Murphy2020} discuss pulsation in a model of a
star in which convection is suppressed by a strong magnetic field.  These 
calculations predict that high-order g modes typical of $\gamma$~Dor 
pulsations are strongly suppressed in Ap stars.  In fact, most of the
199 pulsating Ap stars observed by {\em TESS} are either $\gamma$~Dor stars 
(frequencies not exceeding 5\,d$^{-1}$) or contain multiple low frequencies in 
the $\gamma$~Dor range if they are $\delta$~Sct variables.  Evidently, the 
models of \citet{Saio2005} and \citet{Murphy2020}, as they stand, are not 
supported by observations.

Fig.\,\ref{frqdis} shows the distribution of extracted frequencies in 12544 
non-Ap and 255 pulsating Ap stars.  Note that this is not the distribution of 
stars.  Combination frequencies and harmonics have been excluded.
It can be seen that in both cases there is a continuous distribution 
with a long high-frequency tail. The incidence of high frequencies is
clearly much higher in ``roAp'' than in normal $\delta$~Sct stars.  This does
constitute a distinct difference, but it is not directly related to the
pulsation mechanism.  It would appear that for some reason high frequencies
are preferentially selected in stars with strong magnetic fields.

Unfortunately, we have no idea what drives high frequencies in $\delta$~Sct 
stars, both chemically normal and chemically peculiar.  As already mentioned, 
\citet{Xiong2016} have show that it is possible to drive high frequencies in 
models of normal intermediate mass stars, but only for stars in a small
region of the instability strip.  Without a model, further progress in 
understanding the $\delta$~Sct stars cannot be made. 

One of the most striking and puzzling results of the {\em Kepler} and {\em TESS} 
missions is that the distribution of frequency peaks in $\delta$~Sct stars is 
completely unpredictable.  Two stars with the same effective temperature, 
luminosity, metallicity and rotational velocity should have nearly the same 
observed frequencies.  This turns out not to be the case: the frequencies that 
are observed may be completely different in the two stars \citep{Balona2021c}.
It seems that a highly nonlinear mode selection process is active. 

There is a further discovery which might impact on the driving of the
oscillations.  {\em Kepler} and {\em TESS} light curves indicate that
the light from A and B stars are modulated by rotation \citep{Balona2019c,
Balona2021d}.  Flares are sometimes seen in A and late B stars 
\citep{Balona2012c,Balona2013c, Balona2015a, Balona2021b}.  At this stage, 
there is no model which can account for these phenomena in hot stars.  It is 
also possible that accretion of interstellar material may affect the outer 
layers of A stars, as suggested by \citet{BohmVitense2006}.

Current pulsation models assume that the outer layers are static and devoid of 
magnetic fields.  The treatment of the stellar outer layers in current models 
is therefore incomplete. This may have an impact on the predicted pulsational 
stability and may play an important role in mode selection.  For further 
progress, it is essential to study the atmospheres of A and B stars in order 
to understand the nature of the rotational modulation and to detect possible 
mass motions.  Such a study may lead to pulsation models which better
reproduce the observations.

\section*{Data availability}

The data underlying this article are available in the article.

\section*{Acknowledgments}

I thank the National Research Foundation of South Africa for financial support
and Dr Gerald Handler for useful comments.

This paper includes data collected by the {\it TESS} mission. Funding for the 
{\it TESS} mission is provided by the NASA Explorer Program. Funding for the 
{\it TESS} Asteroseismic Science Operations Centre is provided by the Danish 
National Research Foundation (Grant agreement no.: DNRF106), ESA PRODEX
(PEA 4000119301) and Stellar Astrophysics Centre (SAC) at Aarhus University. 
We thank the {\it TESS} and TASC/TASOC teams for their support of the present
work.

This work has made use of data from the European Space Agency (ESA) mission 
Gaia (\url{https://www.cosmos.esa.int/gaia}), processed by the Gaia Data 
Processing and Analysis Consortium (DPAC,\\
\url{https://www.cosmos.esa.int/web/gaia/dpac/consortium}).\\ 
Funding for the DPAC has been provided by national institutions, in particular 
the  institutions participating in the Gaia Multilateral Agreement.  

This research has made use of the SIMBAD database, operated at CDS, 
Strasbourg, France.  This research has made use of the VizieR catalogue access 
tool, CDS, Strasbourg, France (DOI: 10.26093/cds/vizier). The original 
description of the VizieR service was published in A\&AS 143, 23.

The data presented in this paper were obtained from the Mikulski Archive for 
Space Telescopes (MAST).  STScI is operated by the Association of Universities
for Research in Astronomy, Inc., under NASA contract NAS5-2655.

\bibliographystyle{mn2e}
\bibliography{rotess}

\begin{thebibliography}{51}
\expandafter\ifx\csname natexlab\endcsname\relax\def\natexlab#1{#1}\fi

\bibitem[{{Antoci} {et~al.}(2014){Antoci}, {Cunha}, {Houdek}, {Kjeldsen},
  {Trampedach}, {Handler}, {L{\"u}ftinger}, {Arentoft}, \&
  {Murphy}}]{Antoci2014}
{Antoci} V., {Cunha} M., {Houdek} G., {et~al.}, 2014, \apj, 796, 118

\bibitem[{{Audard} {et~al.}(1998){Audard}, {Kupka}, {Morel}, {Provost}, \&
  {Weiss}}]{Audard1998}
{Audard} N., {Kupka} F., {Morel} P., {Provost} J., {Weiss} W.~W., 1998, \aap,
  335, 954

\bibitem[{{Balmforth} {et~al.}(2001){Balmforth}, {Cunha}, {Dolez}, {Gough}, \&
  {Vauclair}}]{Balmforth2001}
{Balmforth} N.~J., {Cunha} M.~S., {Dolez} N., {Gough} D.~O., {Vauclair} S.,
  2001, \mnras, 323, 362

\bibitem[{{Balona}(2012)}]{Balona2012c}
{Balona} L.~A., 2012, \mnras, 423, 3420

\bibitem[{{Balona}(2013)}]{Balona2013c}
---, 2013, \mnras, 431, 2240

\bibitem[{{Balona}(2014)}]{Balona2014a}
---, 2014, \mnras, 437, 1476

\bibitem[{{Balona}(2015)}]{Balona2015a}
---, 2015, \mnras, 447, 2714

\bibitem[{{Balona}(2018)}]{Balona2018c}
---, 2018, \mnras, 479, 183

\bibitem[{{Balona}(2019)}]{Balona2019c}
---, 2019, \mnras, 490, 2112

\bibitem[{{Balona}(2021{\natexlab{a}})}]{Balona2021b}
---, 2021{\natexlab{a}}, Frontiers in Astronomy and Space Sciences, 8, 32

\bibitem[{{Balona}(2021{\natexlab{b}})}]{Balona2021c}
---, 2021{\natexlab{b}}, arXiv e-prints, arXiv:2109.12574

\bibitem[{{Balona} {et~al.}(2019){Balona}, {Holdsworth}, \&
  {Cunha}}]{Balona2019b}
{Balona} L.~A., {Holdsworth} D.~L., {Cunha} M.~S., 2019, \mnras, 487, 2117

\bibitem[{{Balona} \& {Ozuyar}(2021)}]{Balona2021d}
{Balona} L.~A., {Ozuyar} D., 2021, \apj, 921, 5

\bibitem[{{Baran} \& {Koen}(2021)}]{Baran2021}
{Baran} A.~S., {Koen} C., 2021, \actaa, 71, 113

\bibitem[{{Baran} {et~al.}(2015){Baran}, {Koen}, \& {Pokrzywka}}]{Baran2015}
{Baran} A.~S., {Koen} C., {Pokrzywka} B., 2015, \mnras, 448, L16

\bibitem[{{Bedding} {et~al.}(2020){Bedding}, {Murphy}, {Hey}, {Huber}, {Li},
  {Smalley}, {Stello}, {White}, {Ball}, {Chaplin}, {Colman}, {Fuller},
  {Gaidos}, {Harbeck}, {Hermes}, {Holdsworth}, {Li}, {Li}, {Mann}, {Reese},
  {Sekaran}, {Yu}, {Antoci}, {Bergmann}, {Brown}, {Howard}, {Ireland},
  {Isaacson}, {Jenkins}, {Kjeldsen}, {McCully}, {Rabus}, {Rains}, {Ricker},
  {Tinney}, \& {Vanderspek}}]{Bedding2020}
{Bedding} T.~R., {Murphy} S.~J., {Hey} D.~R., {et~al.}, 2020, \nat, 581, 147

\bibitem[{{Bertelli} {et~al.}(2008){Bertelli}, {Girardi}, {Marigo}, \&
  {Nasi}}]{Bertelli2008}
{Bertelli} G., {Girardi} L., {Marigo} P., {Nasi} E., 2008, \aap, 484, 815

\bibitem[{{B{\"o}hm-Vitense}(2006)}]{BohmVitense2006}
{B{\"o}hm-Vitense} E., 2006, \pasp, 118, 419

\bibitem[{{Bowman} \& {Michielsen}(2021)}]{Bowman2021}
{Bowman} D.~M., {Michielsen} M., 2021, arXiv e-prints, arXiv:2109.10776

\bibitem[{{Breger} {et~al.}(1993){Breger}, {Stich}, {Garrido}, {Martin},
  {Jiang}, {Li}, {Hube}, {Ostermann}, {Paparo}, \& {Scheck}}]{Breger1993}
{Breger} M., {Stich} J., {Garrido} R., {et~al.}, 1993, \aap, 271, 482

\bibitem[{{Cox}(1963)}]{Cox1963}
{Cox} J.~P., 1963, \apj, 138, 487

\bibitem[{{Cunha}(2002)}]{Cunha2002}
{Cunha} M.~S., 2002, \mnras, 333, 47

\bibitem[{{Cunha} {et~al.}(2013){Cunha}, {Alentiev}, {Brand{\~a}o}, \&
  {Perraut}}]{Cunha2013}
{Cunha} M.~S., {Alentiev} D., {Brand{\~a}o} I.~M., {Perraut} K., 2013, \mnras

\bibitem[{{Cunha} {et~al.}(2019){Cunha}, {Antoci}, {Holdsworth}, {Kurtz},
  {Balona}, {Bogn{\'a}r}, {Bowman}, {Guo}, {Ko{\l}aczek-Szyma{\'n}ski},
  {Lares-Martiz}, {Paunzen}, {Skarka}, {Smalley}, {S{\'o}dor}, {Kochukhov},
  {Pepper}, {Richey-Yowell}, {Ricker}, {Seager}, {Buzasi}, {Fox-Machado},
  {Hasanzadeh}, {Niemczura}, {Quitral-Manosalva}, {Monteiro}, {Stateva}, {De
  Cat}, {Garc{\'\i}a Hern{\'a}ndez}, {Ghasemi}, {Handler}, {Hey}, {Matthews},
  {Nemec}, {Pascual-Granado}, {Safari}, {Su{\'a}rez}, {Szab{\'o}}, {Tkachenko},
  \& {Weiss}}]{Cunha2019}
{Cunha} M.~S., {Antoci} V., {Holdsworth} D.~L., {et~al.}, 2019, \mnras, 487,
  3523

\bibitem[{{Dziembowski}(1977)}]{Dziembowski1977a}
{Dziembowski} W., 1977, \actaa, 27, 95

\bibitem[{{Gaia Collaboration} {et~al.}(2018){Gaia Collaboration}, {Brown},
  {Vallenari}, {Prusti}, {de Bruijne}, {Babusiaux}, \&
  {Bailer-Jones}}]{Gaia2018}
{Gaia Collaboration}, {Brown} A.~G.~A., {Vallenari} A., {Prusti} T., {de
  Bruijne} J.~H.~J., {Babusiaux} C., {Bailer-Jones} C.~A.~L., 2018, ArXiv
  e-prints

\bibitem[{{Gaia Collaboration} {et~al.}(2016){Gaia Collaboration}, {Prusti},
  {de Bruijne}, {Brown}, {Vallenari}, {Babusiaux}, {Bailer-Jones}, {Bastian},
  {Biermann}, {Evans}, \& et~al.}]{Gaia2016}
{Gaia Collaboration}, {Prusti} T., {de Bruijne} J.~H.~J., {et~al.}, 2016, \aap,
  595, A1

\bibitem[{{Gontcharov}(2017)}]{Gontcharov2017}
{Gontcharov} G.~A., 2017, Astronomy Letters, 43, 472

\bibitem[{{Holdsworth} {et~al.}(2021){Holdsworth}, {Cunha}, {Kurtz}, {Antoci},
  {Hey}, {Bowman}, {Kobzar}, {Buzasi}, {Kochukhov}, {Niemczura}, {Ozuyar},
  {Shi}, {Szab{\'o}}, {Samadi-Ghadim}, {Bogn{\'a}r}, {Fox-Machado}, {Khalack},
  {Lares-Martiz}, {Lovekin}, {Miko{\l}ajczyk}, {Mkrtichian}, {Pascual-Granado},
  {Paunzen}, {Richey-Yowell}, {S{\'o}dor}, {Sikora}, {Yang}, {Brunsden},
  {David-Uraz}, {Derekas}, {Garc{\'\i}a Hern{\'a}ndez}, {Guzik}, {Hatamkhani},
  {Handberg}, {Lambert}, {Lampens}, {Murphy}, {Monier}, {Pollard},
  {Quitral-Manosalva}, {Ram{\'o}n-Ballesta}, {Smalley}, {Stateva}, \&
  {Vanderspek}}]{Holdsworth2021a}
{Holdsworth} D.~L., {Cunha} M.~S., {Kurtz} D.~W., {et~al.}, 2021, \mnras, 506,
  1073

\bibitem[{{Holdsworth} {et~al.}(2014){Holdsworth}, {Smalley}, {Gillon},
  {Clubb}, {Southworth}, {Maxted}, {Anderson}, {Barros}, {Cameron}, {Delrez},
  {Faedi}, {Haswell}, {Hellier}, {Horne}, {Jehin}, {Norton}, {Pollacco},
  {Skillen}, {Smith}, {West}, \& {Wheatley}}]{Holdsworth2014a}
{Holdsworth} D.~L., {Smalley} B., {Gillon} M., {et~al.}, 2014, \mnras, 439,
  2078

\bibitem[{{Jenkins} {et~al.}(2016){Jenkins}, {Twicken}, {McCauliff},
  {Campbell}, {Sanderfer}, {Lung}, {Mansouri-Samani}, {Girouard}, {Tenenbaum},
  {Klaus}, {Smith}, {Caldwell}, {Chacon}, {Henze}, {Heiges}, {Latham},
  {Morgan}, {Swade}, {Rinehart}, \& {Vanderspek}}]{Jenkins2016}
{Jenkins} J.~M., {Twicken} J.~D., {McCauliff} S., {et~al.}, 2016, in \procspie,
  Vol. 9913, Software and Cyberinfrastructure for Astronomy IV, p. 99133E

\bibitem[{{Kjeldsen} \& {Bedding}(1995)}]{Kjeldsen1995a}
{Kjeldsen} H., {Bedding} T.~R., 1995, \aap, 293, 87

\bibitem[{{Kochukhov} \& {Ryabchikova}(2001{\natexlab{a}})}]{Kochukhov2001b}
{Kochukhov} O., {Ryabchikova} T., 2001{\natexlab{a}}, \aap, 377, L22

\bibitem[{{Kochukhov} \& {Ryabchikova}(2001{\natexlab{b}})}]{Kochukhov2001a}
---, 2001{\natexlab{b}}, \aap, 374, 615

\bibitem[{{Kochukhov} {et~al.}(2007){Kochukhov}, {Ryabchikova}, {Weiss},
  {Landstreet}, \& {Lyashko}}]{Kochukhov2007b}
{Kochukhov} O., {Ryabchikova} T., {Weiss} W.~W., {Landstreet} J.~D., {Lyashko}
  D., 2007, \mnras, 376, 651

\bibitem[{{Kurtz}(1978)}]{Kurtz1978b}
{Kurtz} D.~W., 1978, Information Bulletin on Variable Stars, 1436, 1

\bibitem[{{Kurtz}(1982)}]{Kurtz1982}
---, 1982, \mnras, 200, 807

\bibitem[{{Kurtz} \& {Shibahashi}(1986)}]{Kurtz1986}
{Kurtz} D.~W., {Shibahashi} H., 1986, \mnras, 223, 557

\bibitem[{{Michaud} {et~al.}(1976){Michaud}, {Charland}, {Vauclair}, \&
  {Vauclair}}]{Michaud1976}
{Michaud} G., {Charland} Y., {Vauclair} S., {Vauclair} G., 1976, \apj, 210, 447

\bibitem[{{Murphy} {et~al.}(2020){Murphy}, {Saio}, {Takada-Hidai}, {Kurtz},
  {Shibahashi}, {Takata}, \& {Hey}}]{Murphy2020}
{Murphy} S.~J., {Saio} H., {Takada-Hidai} M., {Kurtz} D.~W., {Shibahashi} H.,
  {Takata} M., {Hey} D.~R., 2020, \mnras, 498, 4272

\bibitem[{{Pecaut} \& {Mamajek}(2013)}]{Pecaut2013}
{Pecaut} M.~J., {Mamajek} E.~E., 2013, \apjs, 208, 9

\bibitem[{{Saio}(2005)}]{Saio2005}
{Saio} H., 2005, \mnras, 360, 1022

\bibitem[{{Saio}(2014)}]{Saio2014}
---, 2014, in IAU Symposium, Vol. 301, Precision Asteroseismology, {Guzik}
  J.~A., {Chaplin} W.~J., {Handler} G., {Pigulski} A., eds., pp. 197--204

\bibitem[{{Samus} {et~al.}(2017){Samus}, {Kazarovets}, {Durlevich}, {Kireeva},
  \& {Pastukhova}}]{Samus2017}
{Samus} N.~N., {Kazarovets} E.~V., {Durlevich} O.~V., {Kireeva} N.~N.,
  {Pastukhova} E.~N., 2017, Astronomy Reports, 61, 80

\bibitem[{{Skiff}(2014)}]{Skiff2014}
{Skiff} B.~A., 2014, VizieR Online Data Catalog, 1, 2023

\bibitem[{{Smalley} {et~al.}(2017){Smalley}, {Antoci}, {Holdsworth}, {Kurtz},
  {Murphy}, {De Cat}, {Anderson}, {Catanzaro}, {Cameron}, {Hellier}, {Maxted},
  {Norton}, {Pollacco}, {Ripepi}, {West}, \& {Wheatley}}]{Smalley2017}
{Smalley} B., {Antoci} V., {Holdsworth} D.~L., {et~al.}, 2017, \mnras, 465,
  2662

\bibitem[{{Smalley} {et~al.}(2015){Smalley}, {Niemczura}, {Murphy}, {Lehmann},
  {Kurtz}, {Holdsworth}, {Cunha}, {Balona}, {Briquet}, {Bruntt}, {De Cat},
  {Lampens}, {Thygesen}, \& {Uytterhoeven}}]{Smalley2015}
{Smalley} B., {Niemczura} E., {Murphy} S.~J., {et~al.}, 2015, \mnras, 452, 3334

\bibitem[{{Soubiran} {et~al.}(2016){Soubiran}, {Le Campion}, {Brouillet}, \&
  {Chemin}}]{Soubiran2016}
{Soubiran} C., {Le Campion} J.-F., {Brouillet} N., {Chemin} L., 2016, \aap,
  591, A118

\bibitem[{{Stibbs}(1950)}]{Stibbs1950}
{Stibbs} D.~W.~N., 1950, \mnras, 110, 395

\bibitem[{{Wenger} {et~al.}(2000){Wenger}, {Ochsenbein}, {Egret}, {Dubois},
  {Bonnarel}, {Borde}, {Genova}, {Jasniewicz}, {Lalo{\"e}}, {Lesteven}, \&
  {Monier}}]{Wenger2000}
{Wenger} M., {Ochsenbein} F., {Egret} D., {et~al.}, 2000, \aaps, 143, 9

\bibitem[{{Xiong} {et~al.}(2016){Xiong}, {Deng}, {Zhang}, \&
  {Wang}}]{Xiong2016}
{Xiong} D.~R., {Deng} L., {Zhang} C., {Wang} K., 2016, \mnras, 457, 3163

\end{thebibliography}
\label{lastpage}

\appendix
\onecolumn

\renewcommand\thetable{A1}

\begin{table*}
\begin{center}
\centerline{\bf APPENDIX}
\caption{List of known roAp or roA stars.  The TIC number, name, variability type
and characteristic pulsation frequency is given, followed by the effective 
temperature and luminosity.  Where available, the rotation period is given.
The last column is the spectral type.  A colon denotes uncertain values.  The 
variability class is expanded to include the following: r - rotational 
sidelobes present; s - variable amplitudes and/or frequencies; 2h,3h - highest 
harmonic present; l - low frequencies present.  A row in italics indicates that
the star was not observed by {\em TESS}.  The variability type in square 
brackets means that the high frequencies are not detected in the {\em TESS} 
light curve}
\label{known}
\begin{tabular}{>{\rowmac}r>{\rowmac}l>{\rowmac}l>{\rowmac}r>{\rowmac}r>{\rowmac}r>{\rowmac}r>{\rowmac}l}
\hline
\multicolumn{1}{c}{TIC}                         &
\multicolumn{1}{l}{Name}                        &
\multicolumn{1}{l}{Var Type}                    &
\multicolumn{1}{c}{$\nu$}                       &
\multicolumn{1}{c}{$T_{\rm eff}$}               &
\multicolumn{1}{c}{$\log \tfrac{L}{L_\odot}$}   &
\multicolumn{1}{c}{$P_{\rm rot}$}               &
\multicolumn{1}{l}{Sp Type}                     \\
\multicolumn{1}{c}{}                            &
\multicolumn{1}{c}{}                            &
\multicolumn{1}{c}{}                            &
\multicolumn{1}{c}{(d$^{-1}$)}                  &
\multicolumn{1}{c}{(K)}                         &
\multicolumn{1}{c}{(dex)}                       &
\multicolumn{1}{c}{(d)}                         &
\multicolumn{1}{c}{}                            \\
\hline
   3814749 & HD 3748                   & ROT+ROA             & 238.205 &  7945 & 1.05 &    1.689 & A5                       \\
   6118924 & HD 116114                 & ROAP                &  66.461 &  7817 & 1.46 &          & F0VpSrCrEu               \\
  12968953 & HD 217704                 & ROAP                & 115.586 &  7669 & 1.28 &          & ApSrEuCr*                \\
  16596302 & HD 52524                  & ROT+ROA:            &  74.156 &  8733 & 0.97 &    2.532 & A0                       \\
  17676722 & HD 63773                  & ACV+ROAP+r          & 167.749 &  8754 & 1.27 &    1.599 & ApSrEuCr*                \\
  22113439 & TYC 2612-1843-1           & ACV+ROAP            &  62.639 &  7900 & 1.38 &    2.119 & A7m:                     \\
\hline
\end{tabular}
\end{center}
\end{table*}

\begin{table*}
\begin{center}
\contcaption{}
\begin{tabular}{>{\rowmac}r>{\rowmac}l>{\rowmac}l>{\rowmac}r>{\rowmac}r>{\rowmac}r>{\rowmac}r>{\rowmac}l}
\hline
\multicolumn{1}{c}{TIC}                         &
\multicolumn{1}{l}{Name}                        &
\multicolumn{1}{l}{Var Type}                    &
\multicolumn{1}{c}{$\nu$}                       &
\multicolumn{1}{c}{$T_{\rm eff}$}               &
\multicolumn{1}{c}{$\log \tfrac{L}{L_\odot}$}   &
\multicolumn{1}{c}{$P_{\rm rot}$}               &
\multicolumn{1}{l}{Sp Type}                     \\
\multicolumn{1}{c}{}                            &
\multicolumn{1}{c}{}                            &
\multicolumn{1}{c}{}                            &
\multicolumn{1}{c}{(d$^{-1}$)}                  &
\multicolumn{1}{c}{(K)}                         &
\multicolumn{1}{c}{(dex)}                       &
\multicolumn{1}{c}{(d)}                         &
\multicolumn{1}{c}{}                            \\
\hline
 173372645 & HD 154708                 & ACV+[ROAP]          & 180.403 &  6812 & 0.86 &    5.366 & A2SrEuCr                 \\
 174475885 & HD 61622                  & SXARI+ROAP:         & 229.188 & 11763 & 2.09 &    0.852 & B9Si                     \\
 176516923 & HD 38823                  & ACV+ROAP            & 165.285 &  7551 & 1.13 &    8.572 & A5SrEuCr*                \\
 176941102 & HD 106563                 & DSCT+ROAP           &  65.460 &  8100 & 1.17 &          & A3m:                     \\
 178575480 & HD 55852                  & ACV+ROAP:           & 207.181 &  8018 & 1.10 &    4.775 & A0SrEuCr                 \\
 179193226 & HD 122570                 & DSCT+ROA            &  99.120 &  7550 & 1.42 &          & A3/5III:                 \\
\hline
\end{tabular}
\end{center}
\end{table*}

\begin{table*}
\begin{center}
\contcaption{}
\begin{tabular}{>{\rowmac}r>{\rowmac}l>{\rowmac}l>{\rowmac}r>{\rowmac}r>{\rowmac}r>{\rowmac}r>{\rowmac}l}
\hline
\multicolumn{1}{c}{TIC}                         &
\multicolumn{1}{l}{Name}                        &
\multicolumn{1}{l}{Var Type}                    &
\multicolumn{1}{c}{$\nu$}                       &
\multicolumn{1}{c}{$T_{\rm eff}$}               &
\multicolumn{1}{c}{$\log \tfrac{L}{L_\odot}$}   &
\multicolumn{1}{c}{$P_{\rm rot}$}               &
\multicolumn{1}{l}{Sp Type}                     \\
\multicolumn{1}{c}{}                            &
\multicolumn{1}{c}{}                            &
\multicolumn{1}{c}{}                            &
\multicolumn{1}{c}{(d$^{-1}$)}                  &
\multicolumn{1}{c}{(K)}                         &
\multicolumn{1}{c}{(dex)}                       &
\multicolumn{1}{c}{(d)}                         &
\multicolumn{1}{c}{}                            \\
\hline
 407929868 & HD 24355                  & ACV+ROAP+r          & 224.294 &  8235 & 1.18 &   27.916 & A5VpSrEu                 \\
\setrow{\itshape}411247704 & HD 196470 & ROAP                & 133.402 &  7279 & 1.15 &          & A2SrEu                   \\
\setrow{\itshape}413938178 & HD 148593 & ROAP                & 134.784 &  7763 & 1.30 &          & A2Sr                     \\
\setrow{\itshape}420687462 & HD 122970 & ACV+ROAP+s          & 129.816 &  6766 & 0.84 &    3.877 & F0CrEuSr       \clearrow \\
 431380369 & HD 20880                  & ACV+ROAP+r+s        &  74.350 &  8629 & 1.66 &    5.197 & ApSr(EuCr)               \\
\setrow{\itshape}432223926 & HD 134214 & ACV+ROAP+2h+s       & 254.621 &  7403 & 0.94 &  248.000 & F2VpSrCrEu     \clearrow \\
\hline
\end{tabular}
\end{center}
\end{table*}

\renewcommand\thetable{A2}

\begin{table*}
\begin{center}
\caption{Chemically peculiar stars known to be $\delta$~Scuti or
$\gamma$~Doradus variables.  The columns are the same as Table\,\ref{known}
(the characteristic frequency is omitted).}
\label{droap}
\begin{tabular}{>{\rowmac}r>{\rowmac}l>{\rowmac}l>{\rowmac}r>{\rowmac}r>{\rowmac}r>{\rowmac}l}
\hline
\multicolumn{1}{c}{TIC}                         &
\multicolumn{1}{l}{Name}                        &
\multicolumn{1}{l}{Var Type}                    &
\multicolumn{1}{c}{$T_{\rm eff}$}               &
\multicolumn{1}{c}{$\log \tfrac{L}{L_\odot}$}   &
\multicolumn{1}{c}{$P_{\rm rot}$}               &
\multicolumn{1}{l}{Sp Type}                     \\
\multicolumn{1}{c}{}                            &
\multicolumn{1}{c}{}                            &
\multicolumn{1}{c}{}                            &
\multicolumn{1}{c}{(K)}                         &
\multicolumn{1}{c}{(dex)}                       &
\multicolumn{1}{c}{(d)}                         &
\multicolumn{1}{c}{}                            \\
\hline
   2849758 & HD 243007                 & DSCT                            &  9033 & 1.99 &          & A1Si                     \\
   3373254 & HD 244248                 & DSCT                            &  7790 & 1.75 &          & A5SiSr                   \\
   3542929 & HD 244372                 & DSCT                            &  8500 & 1.40 &          & A1Sr                     \\
   9211526 & HD 221226                 & GDOR                            &  6840 & 1.12 &          & F3Sr?                    \\
\setrow{\itshape}  9668192 & HD 143517  & DSCT                            &  9150 & 1.22 &          & A3Sr           \clearrow \\
  11646496 & HD 8717                   & DSCT+ACV                        &  7245 & 1.21 &    2.178 & A5pCr                    \\
\hline
\end{tabular}
\end{center}
\end{table*}

\begin{table*}
\begin{center}
\contcaption{}
\begin{tabular}{>{\rowmac}r>{\rowmac}l>{\rowmac}l>{\rowmac}r>{\rowmac}r>{\rowmac}r>{\rowmac}l}
\hline
\multicolumn{1}{c}{TIC}                         &
\multicolumn{1}{l}{Name}                        &
\multicolumn{1}{l}{Var Type}                    &
\multicolumn{1}{c}{$T_{\rm eff}$}               &
\multicolumn{1}{c}{$\log \tfrac{L}{L_\odot}$}   &
\multicolumn{1}{c}{$P_{\rm rot}$}               &
\multicolumn{1}{l}{Sp Type}                     \\
\multicolumn{1}{c}{}                            &
\multicolumn{1}{c}{}                            &
\multicolumn{1}{c}{}                            &
\multicolumn{1}{c}{(K)}                         &
\multicolumn{1}{c}{(dex)}                       &
\multicolumn{1}{c}{(d)}                         &
\multicolumn{1}{c}{}                            \\
\hline
 120598737 & HD 213421                 & EA+DSCT                         &  8716 & 1.80 &          & A0Si?                    \\
 122932611 & HD 134185                 & DSCT                            &  7400 & 1.72 &          & F0Si                     \\
 123231021 & KIC 7582608               & GDOR+ACV:+ROAP                  &  7801 & 1.38 &    9.906 & A7pEuCr                  \\
 123587926 & HD 50285                  & DSCT+ACV                        &  9225 & 2.34 &    1.287 & A0Si                     \\
 127755832 & HD 243356                 & DSCT                            &  9170 & 1.45 &          & A0Si                     \\
 127846482 & HD 243543                 & DSCT                            &  9170 & 1.69 &          & A0Sr                     \\
\hline
\end{tabular}
\end{center}
\end{table*}

\begin{table*}
\begin{center}
\contcaption{}
\begin{tabular}{>{\rowmac}r>{\rowmac}l>{\rowmac}l>{\rowmac}r>{\rowmac}r>{\rowmac}r>{\rowmac}l}
\hline
\multicolumn{1}{c}{TIC}                         &
\multicolumn{1}{l}{Name}                        &
\multicolumn{1}{l}{Var Type}                    &
\multicolumn{1}{c}{$T_{\rm eff}$}               &
\multicolumn{1}{c}{$\log \tfrac{L}{L_\odot}$}   &
\multicolumn{1}{c}{$P_{\rm rot}$}               &
\multicolumn{1}{l}{Sp Type}                     \\
\multicolumn{1}{c}{}                            &
\multicolumn{1}{c}{}                            &
\multicolumn{1}{c}{}                            &
\multicolumn{1}{c}{(K)}                         &
\multicolumn{1}{c}{(dex)}                       &
\multicolumn{1}{c}{(d)}                         &
\multicolumn{1}{c}{}                            \\
\hline
 257921991 & HD 10783                  & SXARI+MAIA                      & 10772 & 1.98 &    4.132 & A2pSiSrCrFe(EuGd)        \\
 260204804 & HD 131910                 & DSCT                            &  6966 & 1.59 &          & F4EuCr                   \\
 267021905 & HD 220071                 & DSCT+ELL                        &  7400 & 1.03 &          & A7/F0V:p:Si:Sr?          \\
 271057422 & HD 75425                  & DSCT                            &  8042 & 1.45 &          & A0SrEuCr?                \\
 276300910 & HD 134799                 & DSCT+ACV+ROAP                   &  9743 & 1.50 &    7.270 & A7SrCrMg                 \\
 283613369 & DM BD+35 4340             & DSCT                            &  7480 & 1.99 &          & F0SiSr?                  \\
\hline
\end{tabular}
\end{center}
\end{table*}

\begin{table*}
\begin{center}
\contcaption{}
\begin{tabular}{>{\rowmac}r>{\rowmac}l>{\rowmac}l>{\rowmac}r>{\rowmac}r>{\rowmac}r>{\rowmac}l}
\hline
\multicolumn{1}{c}{TIC}                         &
\multicolumn{1}{l}{Name}                        &
\multicolumn{1}{l}{Var Type}                    &
\multicolumn{1}{c}{$T_{\rm eff}$}               &
\multicolumn{1}{c}{$\log \tfrac{L}{L_\odot}$}   &
\multicolumn{1}{c}{$P_{\rm rot}$}               &
\multicolumn{1}{l}{Sp Type}                     \\
\multicolumn{1}{c}{}                            &
\multicolumn{1}{c}{}                            &
\multicolumn{1}{c}{}                            &
\multicolumn{1}{c}{(K)}                         &
\multicolumn{1}{c}{(dex)}                       &
\multicolumn{1}{c}{(d)}                         &
\multicolumn{1}{c}{}                            \\
\hline
 437231105 & HD 95158                  & DSCT                            &  7193 & 1.40 &          & A7SrCr?                  \\
 438694338 & HD 117227                 & DSCTHF                          &  7803 & 1.07 &          & A0CrSr                   \\
 445325141 & HD 59660                  & DSCT+ACV                        &  8251 & 2.30 &    4.283 & A0EuCr?                  \\
 445796153 & HD 34740                  & ACV:+DSCTHF                     &  7351 & 1.16 &    5.576 & A0pSrSi                  \\
 449026092 & HD 125467                 & DSCT+ACV                        &  8000 & 1.68 &    3.891 & ApEuCrSr                 \\
 450091404 & HD 287150                 & DSCT                            &  8169 & 1.72 &          & A3SrCr?                  \\
\hline
\end{tabular}
\end{center}
\end{table*}

\renewcommand\thetable{A3}

\begin{table*}
\begin{center}
\caption{List of new roAp and roA stars.  The columns are the same as in
Table\,\ref{known}.  A colon on the variable type signifies stars with a
signal-to-noise ratio S/N$< 5$.}
\label{new}
\begin{tabular}{rllrrrrl}
\hline
\multicolumn{1}{c}{TIC}                         &
\multicolumn{1}{l}{Name}                        &
\multicolumn{1}{l}{Var Type}                    &
\multicolumn{1}{c}{$\nu$}                       &
\multicolumn{1}{c}{$T_{\rm eff}$}               &
\multicolumn{1}{c}{$\log \tfrac{L}{L_\odot}$}   &
\multicolumn{1}{c}{$P_{\rm rot}$}               &
\multicolumn{1}{l}{Sp Type}                     \\
\multicolumn{1}{c}{}                            &
\multicolumn{1}{c}{}                            &
\multicolumn{1}{c}{}                            &
\multicolumn{1}{c}{(d$^{-1}$)}                  &
\multicolumn{1}{c}{(K)}                         &
\multicolumn{1}{c}{(dex)}                       &
\multicolumn{1}{c}{(d)}                           &
\multicolumn{1}{c}{}                            \\
\hline
   4463975 & TYC 5820-1216-1           & ROA+ROT             & 174.753 &  7629 & 1.05 &    0.514 & F0                       \\
   5370150 & HD 55133                  & ROT+ROA:            & 226.528 &  6927 & 0.45 &    3.663 & F6/8(IV)                 \\
   7647538 & HD 27440                  & ROT+ROA:            & 248.231 &  6497 & 1.06 &    6.024 & F5IV/V                   \\
   9171107 & TYC 2479-429-1            & ROT+ROA             & 132.107 &  6523 & 0.1: &    3.096 &                          \\
  12312526 & HD 291240                 & ROA                 &  66.213 &  8249 & 1.11 &          & A2                       \\
  21024812 & HD 14522                  & ACV+ROAP            &  95.265 &  7942 & 1.46 &   13.158 & A8VSrEuSi                \\
\hline
\end{tabular}
\end{center}
\end{table*}

\begin{table*}
\begin{center}
\contcaption{}
\begin{tabular}{rllrrrrl}
\hline
\multicolumn{1}{c}{TIC}                         &
\multicolumn{1}{l}{Name}                        &
\multicolumn{1}{l}{Var Type}                    &
\multicolumn{1}{c}{$\nu$}                       &
\multicolumn{1}{c}{$T_{\rm eff}$}               &
\multicolumn{1}{c}{$\log \tfrac{L}{L_\odot}$}   &
\multicolumn{1}{c}{$P_{\rm rot}$}               &
\multicolumn{1}{l}{Sp Type}                     \\
\multicolumn{1}{c}{}                            &
\multicolumn{1}{c}{}                            &
\multicolumn{1}{c}{}                            &
\multicolumn{1}{c}{(d$^{-1}$)}                  &
\multicolumn{1}{c}{(K)}                         &
\multicolumn{1}{c}{(dex)}                       &
\multicolumn{1}{c}{(d)}                           &
\multicolumn{1}{c}{}                            \\
\hline
 249648704 & HD 215449                 & ROT+ROA:            & 103.634 &  8223 & 1.04 &    4.405 & A7III                    \\
 255988376 & HD 158952                 & ROT+ROA             &  73.386 &  9180 & 1.23 &   12.500 & A0V                      \\
 257168451 & BD+00 2086                & ROA                 &  63.038 &  8501 & 1.12 &          & A2                       \\
 258626846 & TYC 4451-487-1            & ROT+ROA             &  61.364 &  7522 & 0.96 &    3.584 &                          \\
 259017938 & HD 210684                 & ROT+ROA+r           & 116.825 &  7770 & 1.31 &    5.102 & F0                       \\
 259019958 & HD 130232                 & ROA                 &  80.703 &  8210 & 1.14 &          & A0V                      \\
\hline
\end{tabular}
\end{center}
\end{table*}

\label{lastpage}

\end{document}